\documentclass[twocolumn,showpacs,preprintnumbers,amsmath,amssymb]{revtex4}

\usepackage{graphicx}
\usepackage{dcolumn}
\usepackage{bm}

\begin{document}
\title{Beamstrahlung at the NLC}
\author{ G. Bonvicini and N. Powell \\
             \vspace{0.5cm} 
{\it Wayne State University, Detroit MI 48201}\\
}

\date{\today}

\begin{abstract}
The detection of beamstrahlung visible light, divided in its polarization
components, effectively images the beam-beam collision (BBC). Monitoring
and correction of drifts are reviewed. Monitoring of beam jitter
is also possible. The properties of coherent beamstrahlung in the microwave
part of the spectrum
(and its usage)
are introduced.
\end{abstract}


\maketitle

\section{Introduction.}
There has been a lot of simulation and theoretical work about 
beamstrahlung at future linear colliders over the years, yet that work
only scratches the surface of beamstrahlung phenomenology. Beamstrahlung
is of interest to the particle physicist, who needs to know the energy
distribution of colliding beam particles at collision time (dL/dE), and to the
accelerator physicist who must make the beams collide and then steer the spent
beams out of the Interaction Region.

In a series of recent papers we have made clear that recovering complete
information on low energy beamstrahlung effectively recovers most of the
available information about the BBC in $e^+e^-$ 
colliders$[1-3]$. A large-angle infrared beamstrahlung detector 
is being built for CESR\cite{detgen}.

As remarked in Ref.\cite{luckwald}, there are seven transverse degrees
of freedom ($dof$) in the BBC that may decrease luminosity and need to be 
monitored (Fig. 1). 
In the following
the discussion is restricted, without loss of generality, 
to the four BBC presented in Fig. 2.

 \begin{figure}
 \includegraphics[height=90mm,bb=50 150 530 650]{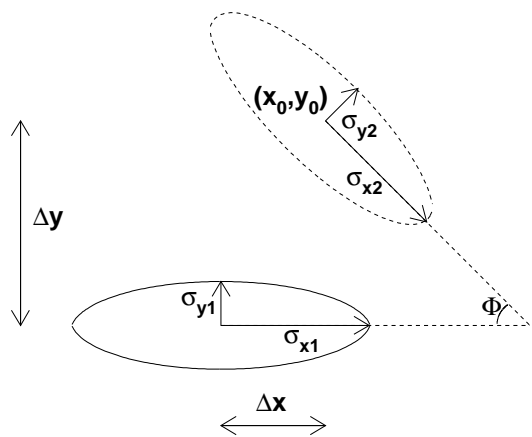}%
 \caption{The seven transverse degrees of freedom in the beam-beam collision.}

\label{fn:ca}
 \end{figure}
 
Without the accurate measurement and monitoring of so many $dof$,  
even the most accurate of 
BBC simulations is of limited usefulness. 
For example, if the bunches collide perfectly (Fig. 2a) the particles
in the center of each beam will have maximal fractional luminosity and 
will produce zero beamstrahlung. If the two beams are offset by 1.5 sigma
(Fig. 2b), the particles in the center of each beam will have close to maximal
fractional luminosity and close to maximal beamstrahlung. The dL/dE curve
is vastly different in the two cases, yet both BBC types contribute usable
amounts of luminosity.

The idea underlying the usage of low energy beamstrahlung 
is actually a simple one. Given the four Maxwell equations,
\begin{eqnarray}
{\bf \nabla \cdot E} &=& 4\pi\rho,\nonumber\\
{\bf \nabla \times B}-{1\over c}{\partial {\bf E}\over \partial t} &=& 
{4\pi\over c}{\bf J},\nonumber\\
{\bf \nabla \times E}+{1\over c}{\partial {\bf B}\over \partial t} 
&=&0,\nonumber\\ 
{\bf \nabla \cdot B} &=& 0,\nonumber
\end{eqnarray}
the beams are the currents, and beamstrahlung is the emitted EM field. 
The equations describe the correlation between currents and fields. We know 
the correlations, and we measure the field to figure out the currents. 
The
fields are vectors, and that is why is necessary to measure their components.
Fig. 3 shows the beamstrahlung polarization components for each beam, 
corresponding to the four BBC sketched in Fig. 2.

In practice, it is difficult to measure the polarization of photons
of energy higher than UV, and that is why we limit ourselves to the study
of low energy beamstrahlung  in the present paper. Coherent
beamstrahlung is available for observation 
in the microwave region, however in that case the polarization information
is not meaningful (see Section IV).

This paper is written with three goals in mind:
\begin{enumerate}
\item to make it clear that large angle incoherent beamstrahlung (IB)
is a necessary feature, 
if polarization information is to be had (Section II). Because large angle
observation is also used for background suppression at CESR, background
suppression at the NLC is also briefly discussed;
\item to discuss the possibility of measuring beam jitter at the NLC (Section III);
\item to introduce coherent beamstrahlung (CB), its properties 
and its potential (Section IV).
\end{enumerate}

\section{Low energy beamstrahlung phenomenology.}

Some points need to be addressed in regard to the usage of this technique
at the NLC. We show that polarization information at the NLC is as 
pristine as at CESR. We also discuss separation of signal and backgrounds
and point out that there are four proposed methods to improve the
S/B ratio.

 \begin{figure}[ht]
 \includegraphics[height=90mm,bb=50 150 530 650]{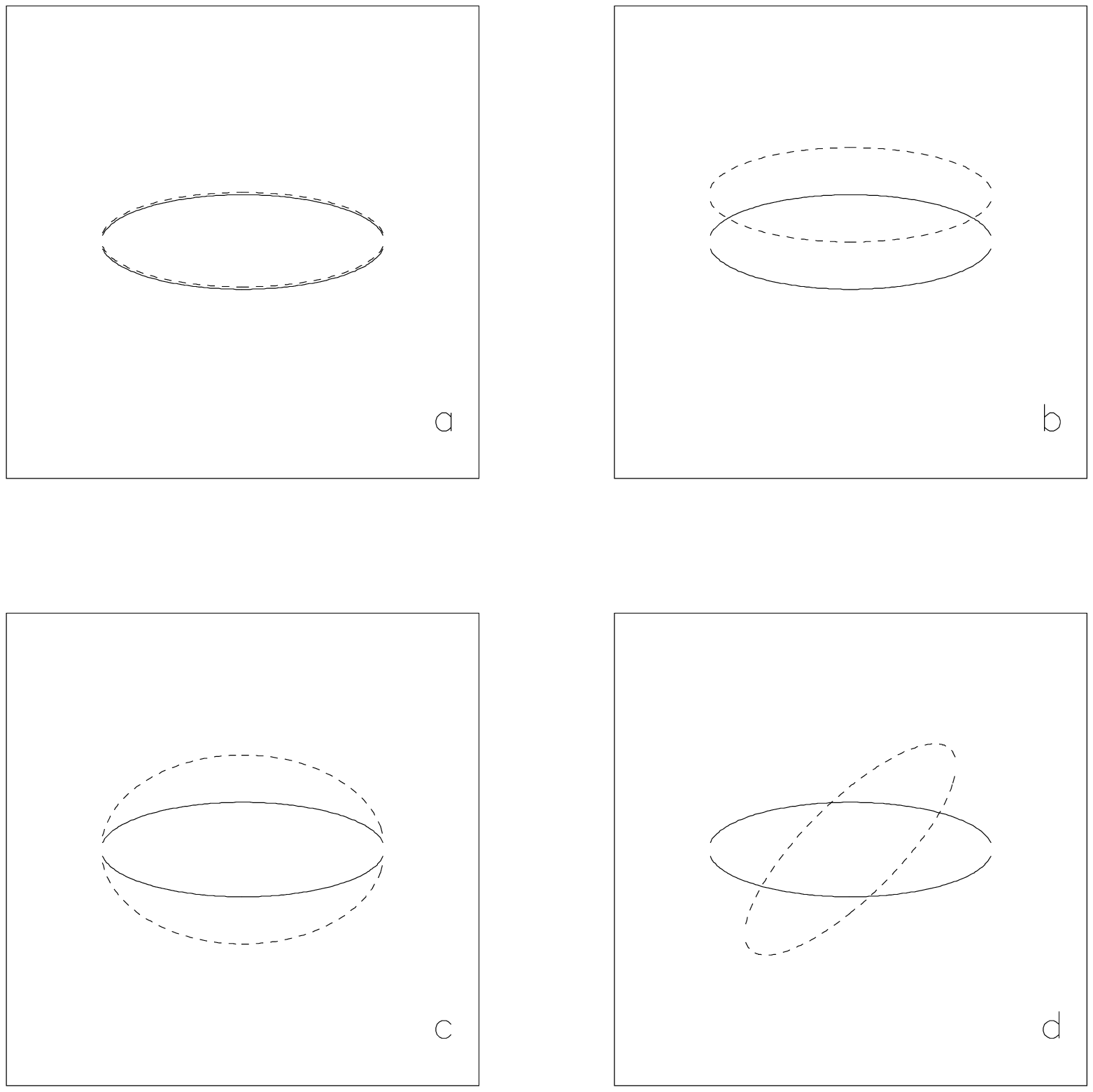}%
 \caption{Three BBCs that lead to wasted
luminosity; a) the beams overlap perfectly, no luminosity is wasted;
b) a $y-$ offset; c) $y-$ bloating;
and d) a beam-beam rotation.
The ``bad'' beam is represented by the dashed ellipse.}

\label{fn:cartoon}
 \end{figure}

Beamstrahlung yields suffer no signficant quantum corrections at low
energies, making classical formulae precise enough to be usable.
The only quantum corrections arise from those beam particles which,
having lost much of their energy through beamstrahlung, have special 
trajectories through the other beam. These corrections (Table I) are at
the percent level and can be ignored throughout this paper.

A single bunch crossing will generate approximately 10$^{12}$
visible photons at the NLC, a good statistics 
to work with. Very low energy photons
have a much larger angular spread than the usual $1/\gamma$ angle.
The total beamstrahlung power at the NLC is of order 1MW, a rate at
which any kind of instrumentation is unlikely to survive. The radiation
is mostly confined to a spot of order 1 mrad.

The Maxwell equations point out the need to measure the polarization of
beamstrahlung. In the classical description of synchrotron radiation (SR),
and therefore of beamstrahlung, the polarization is mostly carried by high
energy photons (of energy comparable to the critical energy)\cite{jackson}.
Even at CESR the polarization of $\approx$10-keV X-rays can not be
measured. At the NLC, a precision measurement of the polarization of 
100-1000 GeV gamma rays is probably out of the question.

Whether one computes visible beamstrahlung by using the classical
formulae\cite{jackson} or by using the ``short-magnet'' 
formulae\cite{coisson},
the polarization content of the radiation, integrated over the solid angle,
is virtually zero (Table I). Ref.\cite{approxi} discusses the 
validity of the two approaches (Ref.\cite{jackson} versus Ref.\cite{coisson})
in different regions of radiation phase space. By applying those formulae,
we find that the ``short magnet'' approximation is most accurate for
all visible beamstrahlung at the NLC, regardless of angle. It is used
to produce the numbers in Table I.
 \begin{figure}[ht]
 \includegraphics[height=90mm,bb=50 150 530 650]{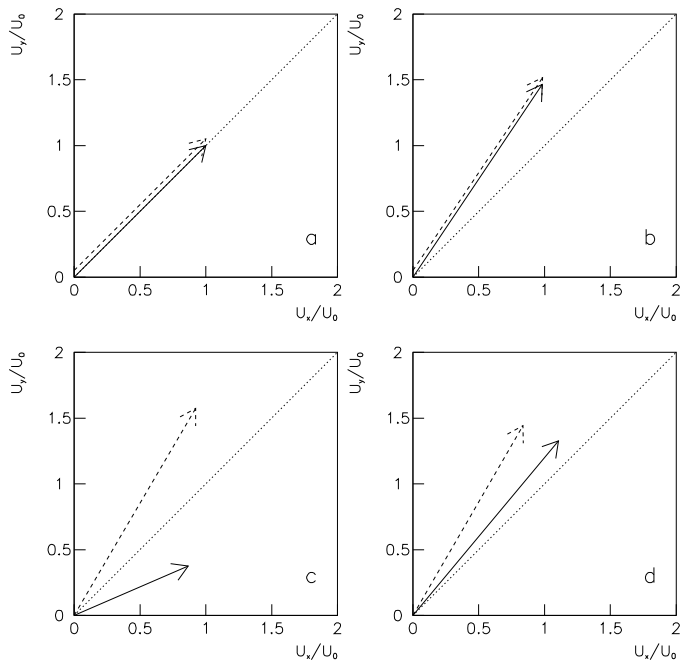}%
 \caption{Beamstrahlung diagrams
corresponding to the four pathologies of Figure~2.
The dashed vectors in parts a) and b) are slightly 
displaced for display purposes.}
\label{fn:cart}
 \end{figure}

\begin{table}[htbp] 
\begin{center}
\begin{tabular}{|c|c|} \hline
 
Beam charge $N$ & $0.75 \times 10^{10}$e \\
Vertical beam width $\sigma_y$  & 3nm \\
Horizontal beam width $\sigma_x$  & 243nm \\
Beam length $\sigma_z$ & 110$\mu m$ \\
Beam energy & 500 GeV \\
\hline
Beamstrahlung average  & 5.4\% \\
energy loss & \\
\hline
Beamstrahlung yield, & 3$\times 10^8$ \\
$350<\lambda<700$nm, $1<\theta<2$ mrad & photons \\
\hline
Beamstrahlung polarization & 0. \\
$350<\lambda<700$nm, $1<\theta<2$ mrad &  \\
\hline
Beamstrahlung yield, offset$=3\sigma_y$ & 5$\times 10^{17}$ \\
$400<\lambda<500 \mu$m, $1<\theta<2$ mrad  &   \\
\hline
Beamstrahlung power, offset$=3\sigma_y$ & 16W  \\
$\lambda>100 \mu$m  &   \\
\hline

\end{tabular}
\end{center}
\caption{NLC nominal parameters and beamstrahlung yield for each bunch 
crossing$[7]$.}
\label{tab:cuts}
\end{table}
However, Ref.\cite{coisson} shows that, if an electron is
subject to a transverse force, its large angle radiation is 
unpolarized as a whole but its azimuthal pattern at large angles exhibits
100\% linear polarization at 8 nodes (for an elementary derivation of
these equations, see Ref.\cite{luckwald}):
\begin{eqnarray}
I_\perp(\theta,\phi,\omega)&=& I_0(\theta,\omega)\cos^2(2\phi),\\
I_\parallel(\theta,\phi,\omega)&=& I_0(\theta,\omega)\sin^2(2\phi),
\end{eqnarray}
where $I_\perp$ and $I_\parallel$ are the polarization 
components w.r.t. to the bending force. $\phi$ is the azimuthal angle with 
respect to the transverse force. At nodes equal to a multiple of
$\pi/4$ the polarization is 100\%, either parallel or perpendicular to
the bending force.

In practice, the beams are 3-dimensional and flat (horizontal transverse size 
much larger than the vertical), with electric charge moving both 
horizontally and vertically as the BBC progresses.
A fixed reference frame has to be chosen (naturally,
one chooses the horizontal and vertical directions) and the 
spacetime charge distributon of the beams always generates  
some particle deflection along each of the axes. Eqs. (1-2) become 
\begin{eqnarray}
I_\perp(\theta,\phi,\omega)&=& I(\theta,\omega)
{U_x\cos^2(2\phi)+U_y\sin^2(2\phi)\over
U_0},\\
I_\parallel(\theta,\phi,\omega)&=& I(\theta,\omega)
{U_x \sin^2(2\phi)+U_y \cos^2(2\phi)\over
U_0}.
\end{eqnarray}

The $U$ 
factors are form factors describing the integral over the
beam charge distribution. The normalizing factor $U_0$ is introduced,
which is the form factor (for either component) when the 
BBC is an exact overlap (Fig. 2a))\cite{luckwald}.
The same factors are the
coordinates of the diagrams in
Fig. 3.

At CESR, 
the detector has been placed at an angle of 10.4 mrad
(or $\sim 100/\gamma$)\cite{detgen}. Above angles $\sim 10/\gamma$,
Eqs. (1-4) hold precisely, so that the polarization components are
disentangled by extracting lights at large angle and 
appropriate azimuthal locations.

In the process, the large angle provides virtually all of the background
suppression, using the fact that a ``short magnet''(the beam) will produce
a radiation cone far wider (in angle) than a
``long magnet'' (the various magnets of CESR)\cite{welch,coisson}.
In short, at CESR the large angle does two things for the experimenter: make 
the polarization observable, and separate signal and background.

At the NLC\cite{nlc}, a beamstrahlung power of order 1MW imposes a stay-clear
cone of 1 mrad (or $\sim 1000/\gamma$).
Clearly at such a large angle the 8-fold pattern of Eqs. (1-4) 
will be available,
however optically one can no longer hope to disentangle signal and background
(assuming a diffraction-limited optical resolution of 1mrad, as in the CESR 
case). Background will be rejected by other methods.

The visible beamstrahlung 
rates at the NLC for each bunch within the train are given in Table I,
assuming full azimuthal acceptance. 
They are certainly abundant and capable of providing subpercent precision
in the measurement of the beamstrahlung diagram.

The principal issue, as usual, is whether the backgrounds can be controlled.
Three background rejection 
methods have already been suggested\cite{gangsun}. 

The first uses the fact that the beamstrahlung pulse is shorter than
the coincident, SR background pulse by a factor of $2\sqrt{2}$. A streak
chamber could disentangle the two components, with a possible background
rejection of order $10^2$.

A second method uses the fact that SR background tends to be strongly ($90\%$)
polarized radially. By extracting only tangential components, one could
reduce backgrounds by one order of magnitude.

A much more powerful method than the previous two was the focus 
of Ref.\cite{gangsun}. An
elliptical grating is the primary mirror, placed so that the Interaction
Point (IP) is located at one of the ellipse foci, and the main collimator
at the other focus. Such a device has extremely shallow field depth ($\sim
100\mu$m at 10 meters distance). The background rejection is
roughly equal to the number of lines in the grating
($\sim 10^4$). It has, however, also a very narrow 
frequency acceptance ($\sim 10^{-5}$), which may prove to be too large
a signal reduction at the NLC. 

Recently, I. Avrutsky\cite{avrutsky} has systematically researched all
possible methods of optical background rejection. He has found that
whole-azimuth imaging, by a means of a ring-like mirror, offers at the same
time a field depth of order one meter and a diffraction-limit of order
0.1 mrad. This method should allow background rejection 
at the $10^{-3}$ level,without any signal bandwidth loss.

\section{Measuring jitter at the NLC.}

The algorithm to make use of these diagrams was worked out 
in Ref.\cite{luckwald}, and is summarized here. A set of four asymmetries,
obtained directly from the diagrams of Fig. 3, 
is defined and ranked,
and the feedback system acts when anyone of the asymmetries becomes 
significantly
different from zero. The wasted luminosity is then expressed as a function
of certain partial derivatives. Tuning of a single corrector magnet (dipole,
quadrupole, or sextupole) will correct any of the ``pathologies'' shown
in Fig.~3. If more than one ``pathology'' exists, it was also proven that 
minimization of the asymmetries the order of their multipole ranking
(dipole first) always converges to the proper overall correction.

The meaning of the diagram is that it can correct the various ways in which
a beam can drift away from its nominal working point 
over time. The diagram both
diagnoses and quantifies drift. It identifies which beam is going bad, which 
corrector magnet needs to be tuned, and by how much it needs to be tuned.
If a machine drifts only (that is, if there is negligible 
beam jitter) 
it is important to notice that after a correction is applied, 
a new diagram is observed. That is why a 4-dimensional diagram monitors a
6-dimensional parameter space. 
The missing $dof$ was identified\cite{luckwald} 
with the smallest of the two $\sigma_y$, 
which can be measured by scanning one beam across the
other (in the process, however, purely passive monitoring is lost).

When measuring jitter, control on the time 
evolution is lost. It is clear that the diagram will still work as 
a monitoring tool. For example, if the beam is oscillating between
the BBC of Fig. 2a) and that of Fig. 2b) the diagram will be oscillating
between those of Fig. 3a) and Fig. 3b). The diagram will
still be able to pick out jitter components that 
other methods can not measure, and the diagram time evolution will
be able to provide information about the frequency, waveform, and amplitude
of the jitter.

There are, however, two limitations that arise in the case of jitter. The 
first and most obvious one is that each singular BBC needs to be recorded.
We have seen in Table I that each BBC provides large statistics, and this
is not expected to be a problem ( for comparison, at CESR rates are integrated
over 1 sec, or 17 millions BBCs). The second limitation is due to lack of
control on the diagram evolution. The diagram jitters uncontrollably and 
the space it covers is equal to its own dimensionality, or four. Three $dof$
are folded in without possibility of detection. 

\section{Coherent beamstrahlung at the NLC.}
If one wants to monitor the luminosity of a machine, any of several low-$Q^2$
QED processes can be used. For the purpose of discussion, let us consider
$e^+e^-\to e^+e^-\gamma$. Most of these events (above a minimum angle) 
consist of one fermion and 
one photon at low angle in the same emisphere, balanced in $p_T$, while
the other fermion continues down the beam pipe. One can reasonably speak
of a radiating beam, the beam in the
same hemisphere as the photon. The event rate in each emisphere, $R_{1,2}$, is
proportional to the luminosity, and therefore to the product of the two beam
populations
\begin{equation}
R_{1,2}\propto L\propto N_1N_2.
\end{equation}

When IB is considered, SR formulae 
are used. In SR theory, the power is
proportional to the beam charge and proportional to the square of the bending
force. This readily translates into a photon rate
\begin{equation}
R_{1}\propto N_1N_2^2.
\end{equation}

When CB is used, the beam moves coherently under the
influence of the EM field of the other beam. Radiation is proportional
to the square of the emitting charge, so that
\begin{equation}
R_{1,2}\propto N_1^2N_2^2.
\end{equation}

A brilliant description of the coherent and incoherent limits for SR 
can be found in Ref.\cite{panofsky}, which concisely derives the
$N_1$ and $N_1^2$ factors in Eqs. (6-7).

Equations (5-7) show at a glance why beamstrahlung is preferrable
to quantum processes - the $N$ factors are huge numbers which make for
more abundant, more precisely measured rates. 
As we will find in this Section, CB has other unique properties.

We have already noted at CESR that the overall $N^3$ dependence
of IB does not favor the early development of the
detector. Weak beams (a factor of ten below nominal) will result in a signal
a thousand times weaker than nominal (at CESR, such a factor is enough to
lower the signal down to the observed background rate).
At the NLC there will be an extensive initial phase of machine
development, with weak, relatively broad beams. CB provides
the large enhancement needed to observe precisely such weak beams.
This is a first, important property of coherent beamstrahlung.

The relativistically invariant coherence condition is 
\[\int d^4x \rho(x) e^{ik\cdot x}\sim 1,\]
where $k$ is the observed 
photon 4-momentum, and $\rho(x)$ is the electron probability
distribution in space-time (normalized to 1). Given that both the beam and the
emitted photons are extremely longitudinal, the coherence condition becomes
simply
\[R= {\lambda\over \sigma_z}\sim 1.\]

If the wavelength is of order of the beam length, the radiation
emitted will be coherent, otherwise it will be incoherent. At CESR, this 
translates to wavelengths greater than 1cm. At the NLC, the wavelengths
of interest are those in excess of 0.1mm. The expected enhancement is also
huge, of order $N_{1,2}\sim 10^{10}$. A second important property of CB
is that one can reasonably expect it to be background-free.
The SR from the magnets will be incoherent, and therefore much weaker, 
because the magnets are
much longer than $\lambda$.

It becomes immediately clear that
at CESR observation is hampered by having a beam pipe whose diameter is
comparable to the wavelength (3cm), resulting in the well-known 
absorption of EM waves as they travel down the beam pipe\cite{jackson}. 
At the NLC, however,
0.1 mm waves are a factor of 25 shorter than the beam pipe diameter and will
be able to travel long distances. Detection of signals is also
much easier in the 0.1 mm range than in the 1cm range. A third important 
property is that at the NLC the experimental conditions are much more 
favorable than at current storage rings (basically due to much shorter
beams).

Another feature of coherent beamstrahlung radiation can be inferred 
directly from 
Fig. 2. It is clear that, for radiation to become coherent, the whole beam
has to move in a certain direction coherently. In Figs. 2c) and 2d), 
different parts of the beams move in opposite directions and interfere
destructively.
It is only in the case of a beam-beam offset (Fig. 2b)) that the beam
as a whole moves vertically. Therefore coherent radiation will only appear
in the presence of a non-zero offset and will primarily measure an offset.
It is also immediately evident that (as long as the offset along the $x-$axis
is not significant)
only the $y-$component of the polarization
will be coherent, as the coherent motion is purely along that direction. Thus
coherent radiation will isolate and amplify two single components (one for
each beam) of the diagrams of Fig. 3. 

To produce quantitative results, the beam-beam simulation program
of Ref.\cite{luckwald} was developed further to include coherent radiation
scoring. This program is one of many cloud-in-cell programs existing on
the market, and since it was developed for CESR, beamstrahlung energy
loss by the beam particles is not included. This is a small deficiency
of the program that does not affect the main results produced below - at the
NLC (Table I) 
the typical beamstrahlung loss is of order a few to several percent
(small corrections like these can be introduced at a later stage). 

When scoring incoherent beamstrahlung, under the assumption discussed above
that one can recover 100\% linear polarization,
one makes use of the following formulae\cite{luckwald} to find the force
exerted by all cells (index $i$) in beam 2 on a cell in the  
beam 1 (index $j$) is
\begin{eqnarray}  
\Delta{\bf r'}_{1j}&=& -{2N_2r_e\over\gamma}\sum {P_{2i}{\bf b}_{ij}
\over b_{ij}^2},\\
{\bf F}_{1j}&=&{\gamma m c^2\over 2\Delta_z}\Delta{\bf r'}_{1j}.
\end{eqnarray}
$\gamma$ is the relativistic factor, $m$ the electron mass, $c$ the speed of
light, $\Delta_z$ the step along the beam collision axis, and $\Delta{\bf r'}$
the (transverse) deflection during such a step. ${\bf F}_{1j}$ 
is the force exerted
on one particle of beam 1 by the whole beam 2.

The $P$ are the fractional charge population in each cell, and {\bf b} is the
transverse impact parameter between the centers of the two cells. 
The energy vector ${\bf U_1}$ for beam 1 is computed by summing

\begin{eqnarray}
U_{1x} = \sum\Delta U_{1xj}&=&{2Nr_e\Delta_z\gamma^2\over 3mc^2}\sum P_j F_{1jx}^2, \\
U_{1y} = \sum\Delta U_{1yj}&=&{2Nr_e\Delta_z\gamma^2\over 3mc^2}\sum P_j F_{1jy}^2.
\end{eqnarray}

The $U$ 
quantities are the low energy power emitted by the beams 
(appropriately scaled by the perfect collision power 
$U_0$, they form the diagrams of Fig. 3). 

When scoring coherent beamstrahlung, the formulae become
\begin{eqnarray}
W_{1x} &=&{2N^2 r_e\Delta_z\gamma^2\over 3mc^2}(\sum 
P_j e^{ik\cdot x} F_{1jx})^2, \\
W_{1y} &=&{2N^2 r_e\Delta_z\gamma^2\over 3mc^2}(\sum 
P_j e^{ik\cdot x} F_{1jy})^2,
\end{eqnarray}
and we apply the normalization condition $W_0=U_0$.
The limitation of the method is that there exists a transition region between
incoherent and coherent beamstrahlung, where the program will not work. The 
two sets of formulae do coincide, in the limit of very large statistics and
short wavelength, and in a way that is consistent with 
Ref.\cite{panofsky}.
However the cell population $C$ inside each beam is finite
(typically $C\sim 3\times10^4$), and we found that the CB program would be
numerically stable only if the coherent enhancement was greater than the
number of cells
\[ C<{U_{CB}\over U_{IB}}.\] 
\begin{figure}[ht]
 \includegraphics[height=90mm,bb=0 0 530 600]{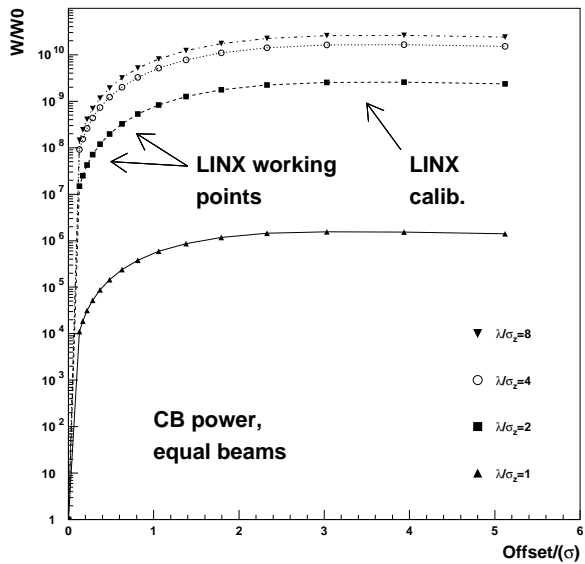}%
 \caption{CB yield as a function of the beam-beam offset. The simulations
were done with NLC nominal conditions (Table I), but weaker beams
($N_1=N_2=0.3\times 10^{10}, \sigma_{y1}=\sigma_{y2}=19$nm). Plots are shown for four different
wavelength-beam length ratios. The LINX working points are discussed in the main text.}

\label{fn:car}
 \end{figure}
\begin{figure}[hb]
 \includegraphics[height=90mm,bb=0 0 530 600]{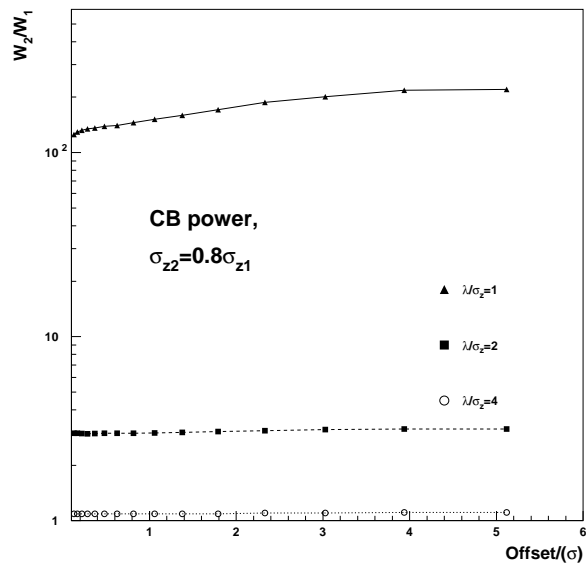}%
 \caption{CB ratio of yields (beam 1 versus beam 2) 
as a function of the beam-beam offset. The simulations
conditions are described in Fig. 4, but $\sigma_{z2}=88\mu$m.}

\label{fn:carto}
 \end{figure}

\begin{figure}[ht]
 \includegraphics[height=90mm,bb=0 0 530 600]{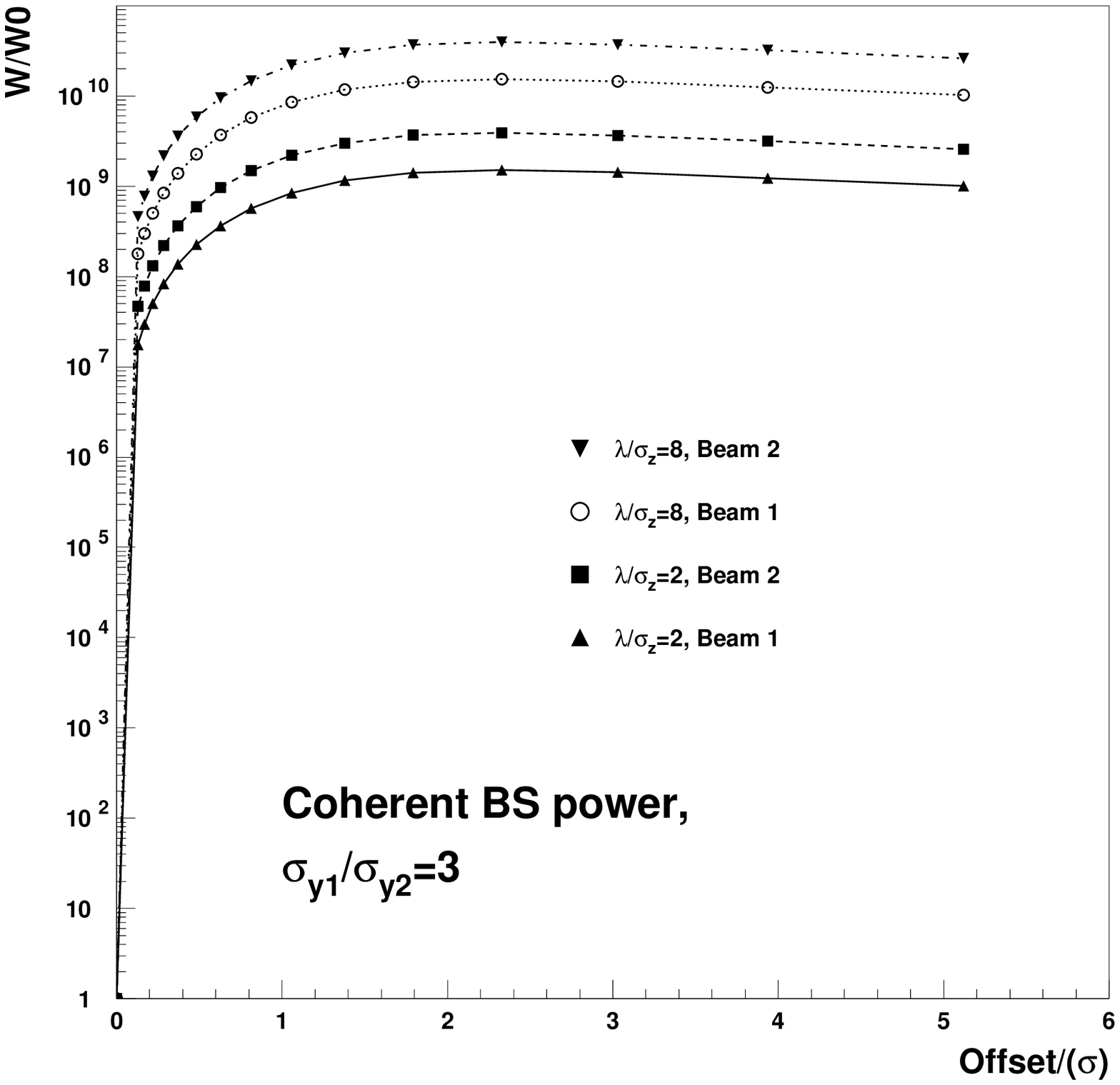}%
 \caption{Same as Fig. 4, but
$\sigma_{y1}=57$nm.}

\label{fn:carto1}
 \end{figure}

The total power emitted in the microwave region may exceed 10W when
full strength NLC beams are offset by a few $\sigma_y$ (Table I). 
The main simulation results are shown in Figs. 4-6 for
``weak'' beams. In Fig. 4, the 
microwave power (in units of IB power) is shown as a function
of the beam-beam vertical offset (in beam width units). The
curves show the dependence of the coherent yield for various R ratios. 

Fig. 5 shows the side-to-side power ratio, for beams which have different beam 
lengths ($\sigma_{z2}/\sigma_{z1})=0.8$. The ordinate
in this plot is the ratio of the powers emitted by the beams. The shorter
beam will obtain coherence at a lower wavelength than the other one,
resulting in substantially more power. From the ratio, and its
dependence on wavelength, one measures accurately the two beam lengths
(with a precision which is probably dominated by uncertainties in the
wavelength in use).

Fig. 6 shows the same plot as Fig. 4, for beams which have different beam 
widths ($\sigma_{y2}/\sigma_{y1}=3$). The slower turn-on of the coherence curve
of the wider beam is noted. Clearly CB measures two distinct degrees of 
freedom, which are, roughly speaking but not exactly, the ratio of the beam-beam offset
and the vertical width of each beam. 

The alert reader will notice that in general the coherent enhancements are
greater than $N$ at large offsets, by a factor of several. 
This is a consequence of the fact that, even in the IB case, the overlap
of one beam's density (peaked at 0) and the other beam's field (peaked
at $\sim 1.6 \sigma_y$) is greater when the two beams are offset, effectively
pushing the enhancement above $N$.

It is clear from Figs. 4 and 6 that the microwave 
power dependence on offset is 
very strong. Together with such large derivatives comes the possibility
to measure jitter to precisions which until today were thought to be 
impossible. As a working example, consider the LINX facility\cite{linx},
which would provide a major proof of principle for future linear colliders.
LINX would produce and collide 50 nm-wide beams, and use them to measure
the beam jitter. A simple way to do that is depicted in Fig. 4. The beams
are brought to a collision, then displaced by a quantity of order 
0.5$\sigma_y$. From Fig. 4, one can see that a 0.1 nm jitter will produce
emitted power fluctuations of order 10\%. Moving the beams to a separation
of 3.0$\sigma_y$ would provide good online 
calibration against possible instrumental jitter.

\section{Conclusions.}

At the NLC, incoherent beamstrahlung (IB) should retain its usefulness
as a near-complete BBC monitor. The expected light signals are large,
and there are, on paper, methods to reduce machine backgrounds.

As beamstrahlung will evolve from CESR to the NLC, it will be required
to do more to monitor the quality and shape of the BBC. At CESR, like at the
NLC, there will be a slow machine drift that incoherent beamstrahlung (IB)
can monitor almost completely by itself (six out of seven $dof$, with 
the seventh one being measurable by scanning beams). 

At NLC, there will be also substantial 
beam jitter (also potentially a 7-dimensional phenomenon). 
IB will be able to monitor only four of these seven $dof$. In part to
counter this limitation, we have introduced the idea of measuring the coherent,
microwave part of beamstrahlung. This part of the spectrum provides
two extra, independent $dof$, bringing the total back to six and effectively
providing almost complete monitoring. CB will also provide precision 
measurements of the beam length, and will work initially with very weak
beams.

Like the beamstrahlung diagrams of Fig. 3, the CB plots presented here are
semi-universal. What that means is that, up to small corrections related
to the beams disruption during the BBC, Figs. (4-6) are universal. That is why
both axes are scaled variables, with each curve depending on a third, scaled
parameter.

By introducing two extra, independent measurements with CB, 
one may wonder whether
the BBC is now fully monitored in a purely passive way, both for drift and 
for jitter. The short answer is
no. Refs.\cite{welch,luckwald} discuss how the total IB power is insensitive
to $\sigma_y$. Consider now a situation where the BBC is jittering
between Fig. 2a) and Fig. 2c). IB will track that jitter, but will only 
provide the time evolution of a quantity which is the ratio of the two 
$\sigma_y$. There will be no CB. The problem will be identified,
and the jitter of the ratio well measured. The absolute size of the smaller
of the two $\sigma_y$ will have to be determined by scanning one beam through
the other. Every other of the seven $dof$ is, however, accounted for and
measurable passively.

Experimental issues will be discussed in a future paper. Once the visible SR 
backgrounds at the IP will be available, one or more of the four background
rejection methods will be implemented in a final design. 

In the case of CB,
the rates are truly enormous (Table I), which allows the usage
of anything sensitive to microwaves (including microantennas inside
the beam pipe). CB will probably be measured above a certain threshold
in the beam-beam offset, of order 0.1$\sigma_y$.
We note that the experimental issues related to CB
are entirely technical, within including how to avoid burning the microwave detector,
how to ensure a very large dynamic range, and how to read the microwave signal
for each bunch (with a time separation of 1.4 nsec). 

Finally, one needs to point out that there are some significant differences
between the low energy beamstrahlung method and the beam-beam deflection method
(see for example\cite{emma}). This method is purely passive, and 
it is sensitive, in a passive way,
to pathologies other than offsets. It also measures the ratio of the two 
$\sigma_y$ in a purely passive way, so that one beam's detuning is diagnosed
instantly. 
When the beams are scanned through one 
another, this method measures both $\sigma_y$ (as opposed to the quadratic
sum of the two). This method will not be affected by different bunch
lengths, and will provide a positive signal when the beams are colliding
properly, whereas the beam-beam deflection provides zero signal.
Finally, beamstrahlung is sensitive to vertical jitter (expressed in units
of $\sigma_y$) far smaller than the beam-beam deflection method.

We thank I. Avrutsky, D. Cinabro and M. Woods for useful discussions.
This work was supported by NSF grants NSF-PHY-0113556, NSF-PHY-0116058 and
NSF-PHY-0101649.

\bibliography{your bib file}

\end{document}